\documentclass[preprint,superscriptaddress,showpacs,preprintnumbers,amsmath,amssymb]{revtex4}
\usepackage{graphicx}
\usepackage{xspace}
\usepackage{epsfig}
\usepackage{multirow}
\usepackage{dcolumn}  
\usepackage{wasysym}
\usepackage[section]{placeins}
\usepackage{subfigure}

\graphicspath{{ps}}

\begin{document}

\vspace*{-3\baselineskip}
\resizebox{!}{3cm}{\includegraphics{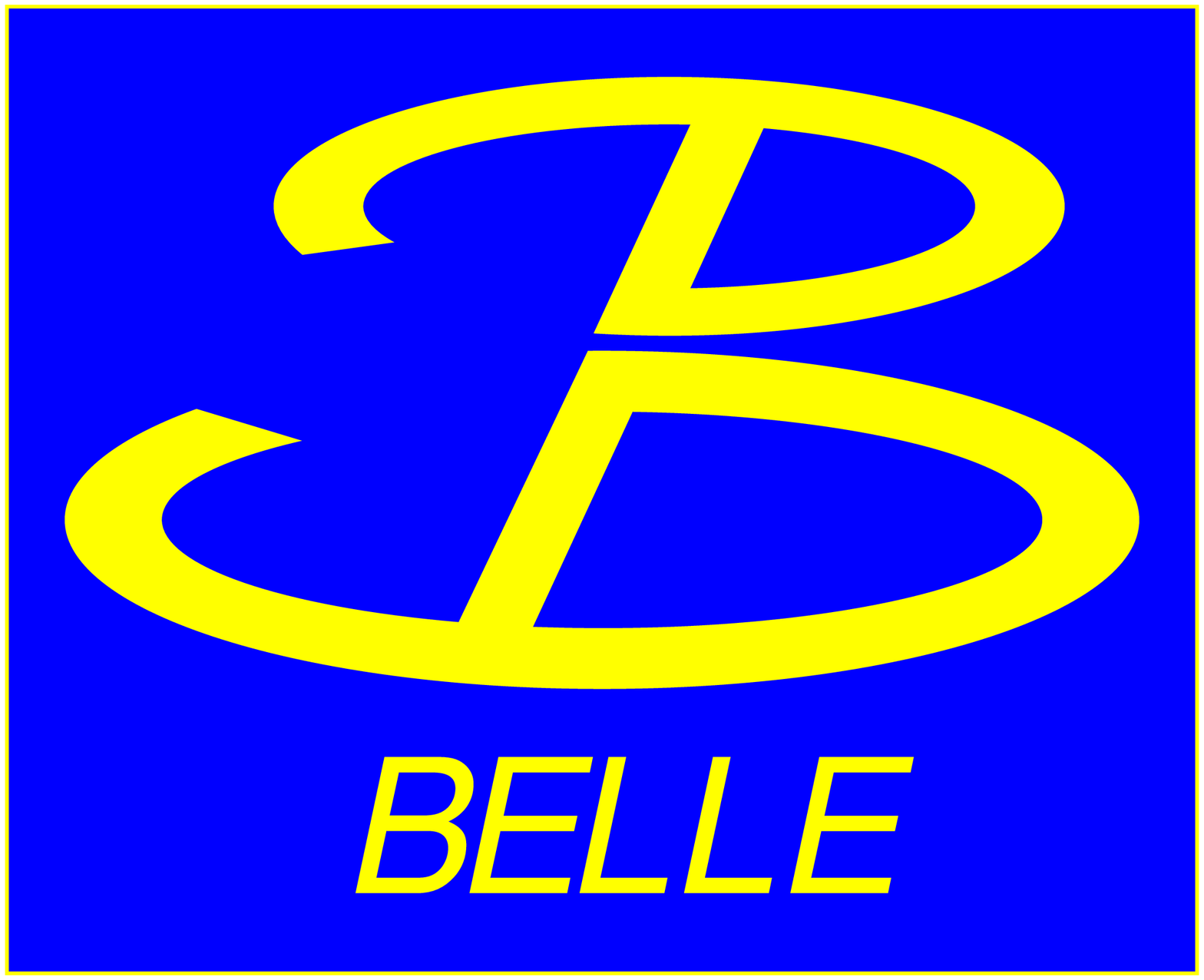}}

\preprint{\vbox{ \hbox{   }
		\hbox{Belle Preprint 2013-22}
		\hbox{KEK Preprint 2013-33}
}}

\title{ \quad\\[1.0cm]\Large \bf \boldmath Search for $B^{0} \to p \bar{\Lambda} \pi^{-} \gamma$ at Belle}


\noaffiliation
\affiliation{University of the Basque Country UPV/EHU, 48080 Bilbao}
\affiliation{Beihang University, Beijing 100191}
\affiliation{University of Bonn, 53115 Bonn}
\affiliation{Budker Institute of Nuclear Physics SB RAS and Novosibirsk State University, Novosibirsk 630090}
\affiliation{Faculty of Mathematics and Physics, Charles University, 121 16 Prague}
\affiliation{University of Cincinnati, Cincinnati, Ohio 45221}
\affiliation{Deutsches Elektronen--Synchrotron, 22607 Hamburg}
\affiliation{Justus-Liebig-Universit\"at Gie\ss{}en, 35392 Gie\ss{}en}
\affiliation{II. Physikalisches Institut, Georg-August-Universit\"at G\"ottingen, 37073 G\"ottingen}
\affiliation{Gyeongsang National University, Chinju 660-701}
\affiliation{Hanyang University, Seoul 133-791}
\affiliation{University of Hawaii, Honolulu, Hawaii 96822}
\affiliation{High Energy Accelerator Research Organization (KEK), Tsukuba 305-0801}
\affiliation{Indian Institute of Technology Guwahati, Assam 781039}
\affiliation{Indian Institute of Technology Madras, Chennai 600036}
\affiliation{Indiana University, Bloomington, Indiana 47408}
\affiliation{Institute of High Energy Physics, Chinese Academy of Sciences, Beijing 100049}
\affiliation{Institute of High Energy Physics, Vienna 1050}
\affiliation{Institute for High Energy Physics, Protvino 142281}
\affiliation{INFN - Sezione di Torino, 10125 Torino}
\affiliation{Institute for Theoretical and Experimental Physics, Moscow 117218}
\affiliation{J. Stefan Institute, 1000 Ljubljana}
\affiliation{Kanagawa University, Yokohama 221-8686}
\affiliation{Institut f\"ur Experimentelle Kernphysik, Karlsruher Institut f\"ur Technologie, 76131 Karlsruhe}
\affiliation{Korea Institute of Science and Technology Information, Daejeon 305-806}
\affiliation{Korea University, Seoul 136-713}
\affiliation{Kyungpook National University, Daegu 702-701}
\affiliation{\'Ecole Polytechnique F\'ed\'erale de Lausanne (EPFL), Lausanne 1015}
\affiliation{Faculty of Mathematics and Physics, University of Ljubljana, 1000 Ljubljana}
\affiliation{Luther College, Decorah, Iowa 52101}
\affiliation{University of Maribor, 2000 Maribor}
\affiliation{Max-Planck-Institut f\"ur Physik, 80805 M\"unchen}
\affiliation{School of Physics, University of Melbourne, Victoria 3010}
\affiliation{Moscow Physical Engineering Institute, Moscow 115409}
\affiliation{Moscow Institute of Physics and Technology, Moscow Region 141700}
\affiliation{Graduate School of Science, Nagoya University, Nagoya 464-8602}
\affiliation{Kobayashi-Maskawa Institute, Nagoya University, Nagoya 464-8602}
\affiliation{Nara Women's University, Nara 630-8506}
\affiliation{National Central University, Chung-li 32054}
\affiliation{National United University, Miao Li 36003}
\affiliation{Department of Physics, National Taiwan University, Taipei 10617}
\affiliation{H. Niewodniczanski Institute of Nuclear Physics, Krakow 31-342}
\affiliation{Nippon Dental University, Niigata 951-8580}
\affiliation{Niigata University, Niigata 950-2181}
\affiliation{University of Nova Gorica, 5000 Nova Gorica}
\affiliation{Osaka City University, Osaka 558-8585}
\affiliation{Pacific Northwest National Laboratory, Richland, Washington 99352}
\affiliation{Panjab University, Chandigarh 160014}
\affiliation{University of Science and Technology of China, Hefei 230026}
\affiliation{Seoul National University, Seoul 151-742}
\affiliation{Sungkyunkwan University, Suwon 440-746}
\affiliation{School of Physics, University of Sydney, NSW 2006}
\affiliation{Tata Institute of Fundamental Research, Mumbai 400005}
\affiliation{Excellence Cluster Universe, Technische Universit\"at M\"unchen, 85748 Garching}
\affiliation{Toho University, Funabashi 274-8510}
\affiliation{Tohoku Gakuin University, Tagajo 985-8537}
\affiliation{Tohoku University, Sendai 980-8578}
\affiliation{Department of Physics, University of Tokyo, Tokyo 113-0033}
\affiliation{Tokyo Institute of Technology, Tokyo 152-8550}
\affiliation{Tokyo Metropolitan University, Tokyo 192-0397}
\affiliation{Tokyo University of Agriculture and Technology, Tokyo 184-8588}
\affiliation{University of Torino, 10124 Torino}
\affiliation{CNP, Virginia Polytechnic Institute and State University, Blacksburg, Virginia 24061}
\affiliation{Wayne State University, Detroit, Michigan 48202}
\affiliation{Yamagata University, Yamagata 990-8560}
\affiliation{Yonsei University, Seoul 120-749}

 \author{Y.-T.~Lai}\affiliation{Department of Physics, National Taiwan University, Taipei 10617} 
 \author{M.-Z.~Wang}\affiliation{Department of Physics, National Taiwan University, Taipei 10617} 
  \author{I.~Adachi}\affiliation{High Energy Accelerator Research Organization (KEK), Tsukuba 305-0801} 
  \author{H.~Aihara}\affiliation{Department of Physics, University of Tokyo, Tokyo 113-0033} 
 \author{D.~M.~Asner}\affiliation{Pacific Northwest National Laboratory, Richland, Washington 99352} 
  \author{V.~Aulchenko}\affiliation{Budker Institute of Nuclear Physics SB RAS and Novosibirsk State University, Novosibirsk 630090} 
  \author{T.~Aushev}\affiliation{Institute for Theoretical and Experimental Physics, Moscow 117218} 
  \author{A.~M.~Bakich}\affiliation{School of Physics, University of Sydney, NSW 2006} 
  \author{A.~Bala}\affiliation{Panjab University, Chandigarh 160014} 
  \author{B.~Bhuyan}\affiliation{Indian Institute of Technology Guwahati, Assam 781039} 
% ...
\collaboration{The Belle Collaboration}


\begin{abstract}
We search for the charmless $B^0$ decay with final state particles $p \bar{\Lambda} 
\pi^{-} \gamma$ using the full data sample that contains $772 \times 10^6 B\bar{B}$ pairs collected at the $\Upsilon(4S)$ resonance with the Belle detector at the KEKB asymmetric-energy $e^+ e^-$ collider. This decay is predicted to 
proceed predominantly via the $b \to s \gamma$ radiative penguin process with a high energy photon. No significant 
signal is found. We set an upper limit of $6.5 \times 10^{-7}$ for the branching fraction 
of $B^{0} \to p \bar{\Lambda} \pi^{-} \gamma$ at
the $90\%$ confidence level.

\end{abstract}

\maketitle
In the Standard Model (SM), the heavy $b$ quark can decay to an energetic $s$ quark and a hard photon via the penguin loop diagram. The inclusive measurement of the branching fraction from $B$ meson decays for 
the above process, ${\mathcal B} ( B \to X_s \gamma)^{\dagger}$~\footnotetext[2]{Throughout this paper, inclusion of charge-conjugate decay modes is always implied.}, is very sensitive to physics beyond the SM since new heavy particles can contribute in the loop at the leading 
order. The up-to-date next-to-next-to-leading order SM calculation 
gives ${\mathcal B} ( B \to X_s \gamma) = (3.15 \pm 0.23) \times 10^{-4}$ for $E_\gamma > 1.6$ 
GeV~\cite{Misiak}, which is consistent with the current world average of the experimental 
results, ${\mathcal B} ( B \to X_s \gamma) = (3.40 \pm 0.21) \times 10^{-4}$~\cite{Lees:2012wg, Limosani:2009qg, Chen:2001fja, PDG}. 

In the Monte Carlo (MC) simulation of the $s \to X_s$ fragmentation and hadronization processes 
by JETSET~\cite{JETSET}, the $X_s$ with a $\Lambda$ in the final state contributes only at the 1\% level. 
This is consistent with the known baryonic $B$ decay rate, ${\mathcal B} ( B^+ \to p \bar\Lambda \gamma) 
= (2.45 ^{+0.44}_{-0.38} \pm 0.22)\times 10^{-6}$~\cite{Wang, PDG}. 
There is an intriguing feature of this three-body decay: the mass of the $p \bar\Lambda$ system is peaked near threshold. A similar feature is seen in many other hadronic three-body $B$ decay processes. In multi-body hadronic baryonic $B$ decays, hierarchy in the branching fractions is also observed; e.g., ${\mathcal B} ( B^+ \to p \bar{\Lambda}\pi^+\pi^-) > {\mathcal B} ( B^0 \to p \bar{\Lambda}\pi^-) > {\mathcal B} ( B^+ \to p \bar{\Lambda})$ and ${\mathcal B} ( B^0 \to p \bar{\Lambda}_c^-\pi^+\pi^-) > {\mathcal B} ( B^+ \to p \bar{\Lambda}_c^-\pi^+) > {\mathcal B} ( B^0 \to p \bar{\Lambda}_c^-)$~\cite{Chen:2009xg, Aubert:2009am, Wang, Tsai:2007pp, Dytman:2002yd, Gabyshev:2002zq, Aubert:2008ax, Abe:2004sr, Gabyshev:2002dt, PDG}.

\begin{figure}[htb]
\centering
\includegraphics[height=4.5cm]{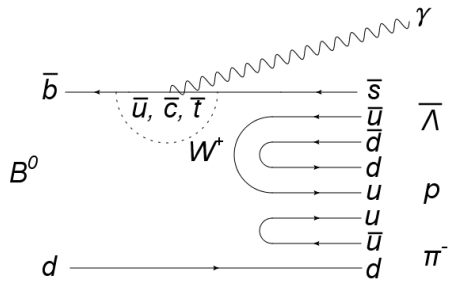}
\caption{Decay diagram of $B^{0} \to p \bar{\Lambda} \pi^{-} \gamma$}.
\label{fg:fenyman} 
\end{figure}

These features motivate our interest in the search for $ B^0 \to p \bar\Lambda \pi^-\gamma$. Figure~\ref{fg:fenyman} shows a possible decay diagram at the
quark level for $B^{0} \to p \bar{\Lambda} \pi^{-} \gamma$. It proceeds via the radiative penguin process.
The $p \bar\Lambda$ system in this decay will have a smaller maximum kinetic energy than in $ B^+ \to p \bar\Lambda \gamma$ 
due to the extra pion in the $X_s$ fragmentation process. 
This matches the threshold enhancement effect naturally and implies a higher decay rate~\cite{hou}.
The measured branching fraction of $B^0 \to p \bar\Lambda \pi^-\gamma$ can be useful to tune the parameters in JETSET and, in the case of a large enhancement of the branching fraction, the uncertainty on the measurement of ${\mathcal B} ( B \to X_s \gamma)$ would be reduced using a sum of exclusive final states.


We use the full data sample (711 $\rm{fb^{-1}}$) that contains $772 \times 10^6 B\bar{B}$ pairs collected at the $\Upsilon(4S)$ resonance with the Belle detector~\cite{Belle} at the KEKB asymmetric-energy $e^+ e^-$ collider~\cite{KEKB} for this search.
The Belle detector is a large-solid-angle magnetic
spectrometer that consists of a silicon vertex detector (SVD),
a 50-layer central drift chamber (CDC), an array of
aerogel threshold Cherenkov counters (ACC), 
a barrel-like arrangement of time-of-flight
scintillation counters (TOF)
and an electromagnetic calorimeter comprised of CsI(Tl) crystals located inside 
a superconducting solenoid that provides a 1.5~T
magnetic field. 
An iron flux-return located outside the solenoid is instrumented to detect $K_L^0$ mesons and to identify muons. 
The detector is described in detail elsewhere~\cite{Belle}.~
The data set used in this analysis was collected with two different inner detector configurations. About $152 \times 10^6 B\bar{B}$ pairs were collected with a beam pipe of radius 2 cm and with 3 layers of SVD, while the rest of the data set was collected with a beam pipe of radius 1.5~cm and 4 layers of SVD \cite{svd2}.

Large MC samples for signal and different backgrounds are generated with EvtGen~\cite{ref:EvtGen} and simulated  
under GEANT3~\cite{geant} with the configuration of the Belle detector.  These samples are used to obtain the expected 
distributions of various physical quantities for signal and background, optimize the selection 
criteria, and determine the signal selection efficiency.

The selection criteria for the final state charged particles 
in $B^{0} \to p \bar\Lambda \pi^-\gamma$ are based on information obtained from
the tracking systems (SVD and CDC) and the hadron identification systems (CDC, ACC, and TOF). 
The proton and pion from $B^0$ decay are required to have a point of 
closest approach to the interaction point (IP) within $\pm 0.3$ cm in the transverse ($x$-$y$) plane, and within $\pm 3$ cm along the $z$ axis, where the +$z$ direction is opposite the positron beam direction. The likelihood values of each track for different particle types, $L_{p}$, $L_{K}$, and $L_{\pi}$, are determined from the information provided by the hadron identification system. The track is identified as a proton 
if $L_p/(L_p+L_K) > 0.6$  and $L_p/(L_p+L_{\pi}) > 0.6$, or as a pion if $L_{\pi}/(L_{\pi}+L_K)>0.6$. The efficiency for identifying a pion is about 95\%, depending on the momentum of the track, while the probability for a kaon to be misidentified as a pion is less than 10\%. The efficiency for identifying a proton is about 95\%, while the probability for a kaon or a pion to be misidentified as a proton is less than 10\%. The efficiency and misidentification probability are averaged over the momentum of the particles in the final state.
We reconstruct a $\Lambda$ candidate from its decay to $p\pi^{-}$. Each $\Lambda$ candidate must have a displaced vertex with its momentum vector being consistent with an origin at the IP. 
The proton from $\Lambda$ decay is required to satisfy the proton criteria described above whereas the pion daughter
has no such requirement. The reconstructed $\Lambda$ mass 
should satisfy 1.111 GeV/$c^{2} < M_{p\pi^{-}} < $1.121 GeV/$c^{2}$, and this constraint retains about 81.7\% of total signal events.
The hard photon must have an energy greater than 1.7 GeV in the center-of-mass (CM) frame.

Candidate $B$ mesons are identified with kinematic variables calculated in the CM 
frame: the beam-energy-constrained mass $M_{\rm bc}\equiv \sqrt{E^{2}_{\mathrm{beam}}/c^{4} - |p_{B}/c|^{2}}$, 
and the energy difference $\Delta E \equiv E_B - E_{\mathrm{beam}}$, where $E_{\mathrm{beam}}$ is the beam energy, 
and $p_B$ and $E_B$ are the momentum and energy of the reconstructed $B$ meson, respectively. 
The candidate region is defined 
as $M_{\rm bc} >$ 5.24 GeV/$c^{2}$ and $-0.4$ GeV $< \Delta E <$ 0.3 GeV, 
and the signal region is defined as $M_{\rm bc} >$ 5.27 GeV/$c^{2}$
and $ |\Delta E| <$ 0.05 GeV.

The dominant background for $B^{0} \to p \bar{\Lambda} \pi^{-} \gamma$ in the candidate 
region is from the continuum $e^{+}e^{-} \to q\bar{q}~(q = u,d,s,c)$ processes. We distinguish 
the jet-like continuum background relative to the more spherical $B\bar{B}$ signal 
using a Fisher discriminant discussed in Ref.~\cite{KSFW}. The Fisher 
discriminant is a linear combination of several shape variables with coefficients that are 
optimized to separate signal and background. An independent variable, cos$\theta_{B}$, where $\theta_{B}$ is the angle between the reconstructed $B$ flight 
direction and the beam direction in the CM frame, is combined with the Fisher discriminant to form signal and background probability density 
functions (PDFs). These PDFs, obtained separately from 
signal and continuum MC simulations, give the event-by-event signal and background 
likelihoods, $\mathcal{L}_{S}$ and $\mathcal{L}_{B}$. We apply a constraint on the 
likelihood ratio,
$\mathcal{R} \equiv \mathcal{L}_{S}/(\mathcal{L}_{S}+\mathcal{L}_{B}) > 0.85$, to suppress the continuum 
background. The value of the $\mathcal{R}$ constraint is determined by maximizing the figure of merit, 
defined as $N_{S}/\sqrt{N_{S}+N_{B}}$, where $N_{S}$ denotes the expected number of signal events 
in the signal region with an assumed branching fraction ($10^{-5}$), 
and $N_{B}$ denotes the expected number of continuum background events in the signal region. The constraint on $\mathcal{R}$ removes 97\% of continuum background while retaining 61\% of the signal.

If more than one $B$ candidate is found in a single event, we choose
the one with the smallest $\chi^{2}_{B}+\chi^{2}_{\Lambda}$ value, using the goodness of fit values $\chi^{2}_{B}$ and
 $\chi^{2}_{\Lambda}$ are $\chi^{2}$ from the $B$ and $\Lambda$ vertex fits, respectively. The vertex fits only use the charged daughters. Multiple candidates are mainly due to the misreconstruction using a pion from the other $B$ meson and are found in 9.8$\%$ of the data; the average multiplicity is 2.15.

Other important backgrounds in the candidate region include 
$B$ decays through the $b \to c$ process (generic $B$ decays), charmless (i.e., `rare`) $B$ decays and the self-crossfeed events. Since the generic $B$ decays do not cause any peaking structure
in the candidate region and their yields are much less than that of continuum background, we merge these with
the continuum background. The remaining backgrounds have a peaking structure in $\Delta E$ and $M_{\rm bc}$, although the overall shapes are 
quite different from the signal shapes. Based on the rare-$B$ MC simulation, the following seven modes are found to contribute to the candidate region: $B^0 \to p \overline{\Lambda} \rho^{-}$, $B^0 \to p \overline{\Sigma^0} \rho^{-}$, $B^0 \to p \overline{\Lambda} \pi^{-} \eta$, $B^+ \to p \overline{\Lambda} \pi^0$, $B^+ \to p \overline{\Sigma^0} \pi^0$, $B^+ \to p \overline{\Lambda} \gamma$, and $B^+ \to p \overline{\Lambda} \eta$. Only two of these, $B^+ \to p \overline{\Lambda} \pi^0$ and $B^+ \to p \overline{\Lambda} \gamma$~\cite{Wang, PDG}, have been measured experimentally.  
For the $B^{0} \to p \bar{\Lambda} \pi^{-} \gamma$ self-crossfeed events, candidate $B$ events are
misreconstructed using a slow pion from the other $B$ meson. 
According to MC simulation, we find 42\% of events are self-crossfeed events and cannot be removed without losing significant signal. We rely on the fitting method to distinguish signal from these backgrounds.

The signal yield of the $B^{0} \to p \bar{\Lambda} \pi^{-} \gamma$ mode is extracted 
from a two-dimensional extended unbinned maximum likelihood fit, with the likelihood defined as 
\begin{equation}
\mathcal{L}=\frac{e^{-\sum_{j}N_{j}}}{N!}\prod^{N}_{i=1}(\sum_{j}N_{j}P_{j}(M_{\rm bc}^{i},\Delta E^{i})),
\end{equation}
where $N$ is the total number of candidate events, $N_{j}$ is the number of  
events in category $j$, and $P_{j}$ represents the  
value of the corresponding two-dimensional PDF, and $M_{\rm bc}^{i}$ ($\Delta E^{i}$) is the $M_{\rm bc}$ ($\Delta E$) value of the $i$-th candidate.  
There are five PDFs in the fit: signal, self-crossfeed, continuum background and the 
two measured rare decay modes ($B^+ \to p \bar{\Lambda} \pi^0$ and $B^+ \to p \bar{\Lambda} \gamma$). The other five rare $B$ modes are considered only in the systematic uncertainties, as discussed later.
We use two-dimensional smoothed histograms to represent the $M_{\rm bc}$-$\Delta E$ PDFs of 
signal, self-crossfeed and two measured rare $B$ modes.
The signal PDF is calibrated by comparing the difference between data and MC simulation for the $B^{+}\to K^{*+}\gamma$ control sample. 
The PDF that describes the continuum background is a product of an ARGUS function~\cite{argus} in $M_{\rm bc}$  
and a second-order polynomial in $\Delta E$.  
The ratio of self-crossfeed events to signal events is fixed, which is estimated from the MC simulation, and the yields 
of two measured rare modes are also fixed according to the measured 
branching fractions~\cite{Wang, PDG}. The free parameters in the fit are the signal yield, the continuum yield and the continuum shape parameters.

\begin{figure}[htb]
\centering
\includegraphics[width=0.48\textwidth]{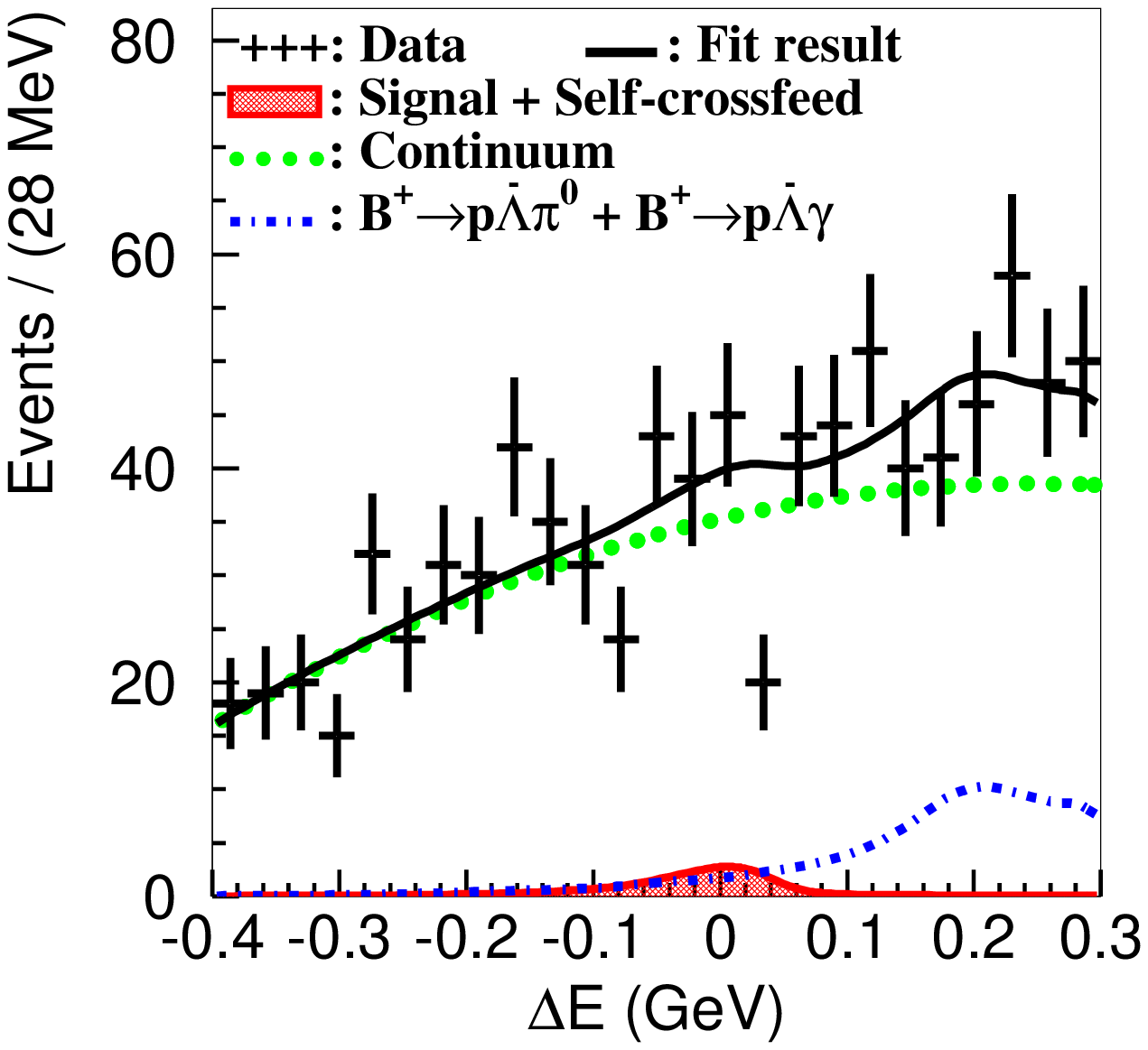}
\includegraphics[width=0.48\textwidth]{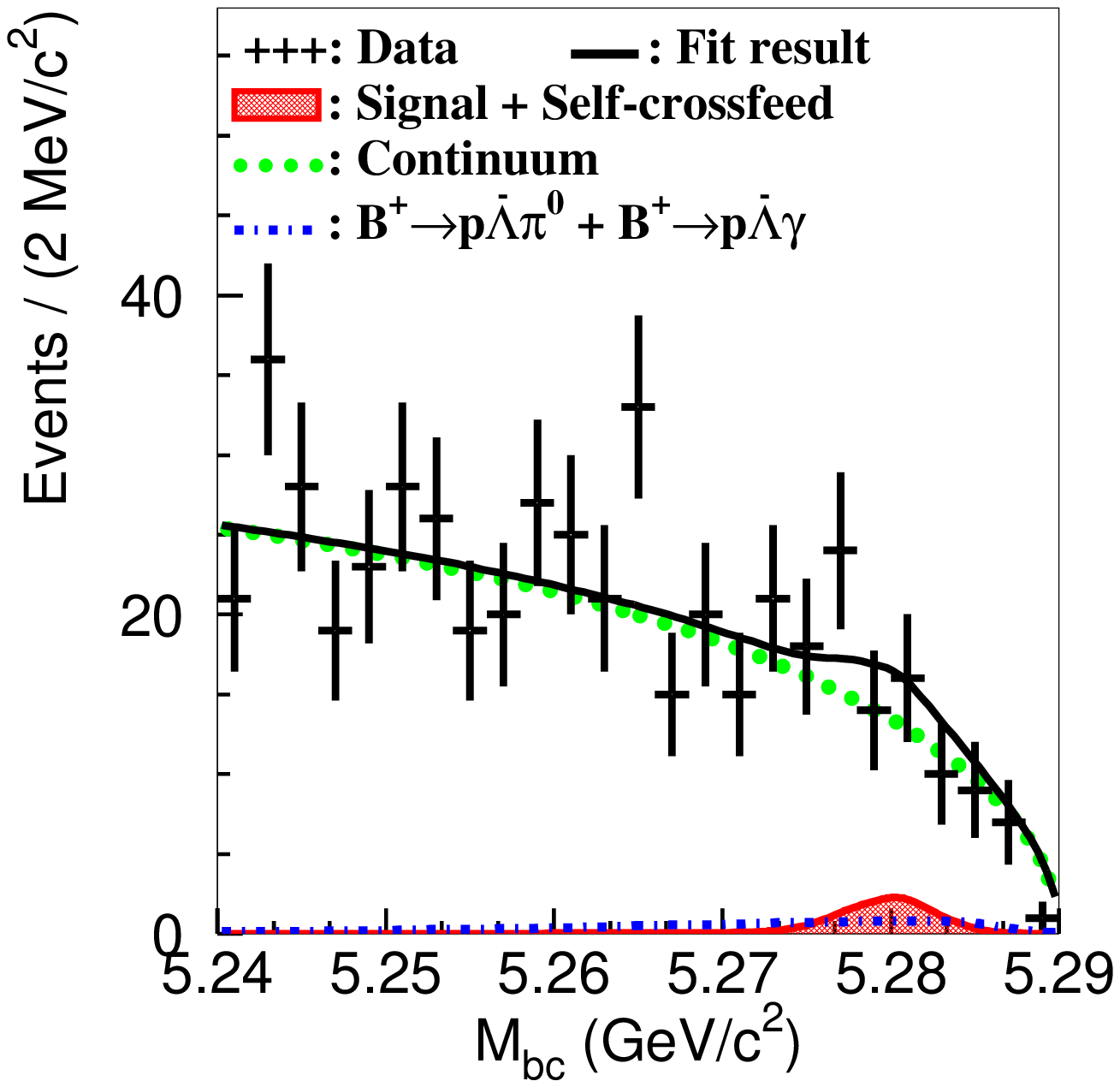}
\caption{Fit results of $B^{0} \to p \bar{\Lambda} \pi^{-} \gamma$. The top plot shows the $\Delta E$ distribution for $M_{\rm bc} >$ 5.27 GeV/$c^{2}$ and the bottom one shows $M_{\rm bc}$ for $|\Delta E| <$ 0.05 GeV.
The points with error bars are data; the solid line is the fit result; the green dotted line is continuum background; 
the blue dash-dotted line is the combination of $B^+ \to p \bar{\Lambda} \pi^0$ 
and $B^+ \to p \bar{\Lambda} \gamma$, and the red area is the combination of signal and self-crossfeed.}
\label{fg:fit} 
\end{figure}

The projections of the fit are shown in Fig.~\ref{fg:fit}. 
The fitted signal yield is $9.5^{+11.5}_{-10.7}$ with a statistical significance of 0.9. The statistical significance 
is defined as $\sqrt{-2\rm{ln}(\mathcal{L}_{0}/\mathcal{L}_{max})}$, where $\mathcal{L}_{0}$ and $\mathcal{L}_{\rm{max}}$ are the likelihood values obtained by the fit with and without the signal yield fixed to zero, respectively. 

The branching fraction is calculated using 
\begin{equation}
\mathcal{B}=\frac{N_{\rm{sig}}}{\epsilon \times N_{B\bar{B}}},
\end{equation}
where $N_{\rm{sig}}$, $N_{B\bar{B}}$, and $\epsilon$ are the fitted signal yield, 
the number of $B\bar{B}$ pairs, and the reconstruction efficiency of signal, respectively. We assume that charged 
and neutral $B\bar{B}$ pairs are produced equally at the $\Upsilon(4S)$. 
We calibrate the reconstruction efficiency estimated using the MC simulation by including in $\epsilon$ a factor $\varepsilon_{\mathcal{R}}\times\varepsilon_{HID}$, where $\varepsilon_{\mathcal{R}}$(= 0.973 $\pm$ 0.018) and $\varepsilon_{HID}$(= 0.928 $\pm$ 0.011) refer to the corrections due to the constraint on $\mathcal{R}$ and the hadron identification, respectively.
Here, $\varepsilon_{\mathcal{R}}$ is obtained from the control sample study of $B^{+}\to K^{*+}\gamma$; $\varepsilon_{HID}$ is determined by various control samples with different particle types such as $\Lambda \to p \pi^-$ and $D^{*+} \to D^{0}\pi^{+}$ with $D^{0} \to K^{-}\pi^{+}$. The calibrated reconstruction efficiency for the signal is $\epsilon=~$5.27\%.

Sources of various systematic uncertainties on the branching fraction calculation are shown in Table~\ref{tb:sys}. The uncertainty due to the total number of $B\bar{B}$ pairs is 1.4\%. The uncertainty due to the charged-track reconstruction efficiency is estimated to be 0.35\% per track by using the partially reconstructed $D^{*+}\to D^0 \pi^{+}$ with $D^0 \to \pi^+ \pi^- K^{0}_{S}$ events. The uncertainty due to $\Lambda$ selection is estimated by a control sample study of $\Lambda \to p\pi^-$. The uncertainty due to photon selection is evaluated with a radiative Bhabha sample to be 2.2\%. The uncertainties due to the $\mathcal{R}$ constraint and the signal PDF shape are estimated using the control sample of $B^{+}\to K^{*+}\gamma$. Because of the presence of the self-crossfeed PDF in the fit, the uncertainty due to the signal PDF shape is inflated by a factor of $\sqrt{2}$. The uncertainty due to the signal decay model is estimated to be 5.1\% by using different decay models. For instance, the base decay model of our study is $B^{0} \to X_{s} \gamma$ with $X_{s} \to p \bar{\Lambda} \pi^{-}$decaying uniformly in phase space; the mass of $X_{s}$ has a simple Breit-Wigner distribution with a mean value at 2.5 GeV/$c^{2}$ and a 0.3 GeV/$c^{2}$ width. An alternate model is $X_{s} \to X_{pl} \pi^{-}$ with $X_{pl} \to p \bar{\Lambda}$, where $X_{pl}$ 
stands for the threshold peak measured in Ref.~\cite{Wang}. The uncertainties for the two measured rare modes discussed above are estimated by varying each yield in the fit by $\pm 1\sigma$, where $\sigma$ denotes the measurement error on the branching fraction. The uncertainty for the five unmeasured rare modes discussed above is estimated by incorporating their PDFs in the fit and floating their yields. As the signal yield is reduced by this fit, we did not include this effect in the upper limit calculation described below to get a conservative upper limit.
The overall systematic uncertainty due to rare $B$ decays is 8.2\% and dominates in this measurement.

\begin{table}[htbp]
\begin{center}
\caption{Summary of the systematic uncertainties (in \%) on the branching fraction.}
\begin{tabular}{c|c}
\hline
\hline
$N_{B\bar{B}}$ & 1.4\\\hline
Tracking & 1.4 (4 tracks)\\\hline
Hadron identification & 0.6 (2 protons)\\
& 1.1 (pion)\\\hline
$\Lambda$ selection & 3.3 \\\hline
Photon selection & 2.2 \\\hline
Reconstruction eff. (MC statistics)& 2.2 \\\hline
$\mathcal{B}(\Lambda \to p \pi^{-})$ & 0.8 \\\hline
$\mathcal{R}$ constraint & 1.9 \\\hline
PDF shape & 4.1 \\\hline
Signal decay model & 5.1 \\\hline
Rare $B$ decays & 8.2 \\\hline
\hline
Total & 11.8\\\hline
\hline
\end{tabular}
\label{tb:sys}
\end{center}
\end{table}

Since the observed yield for $B^{0} \to p \bar{\Lambda} \pi^{-} \gamma$ is not significant, we evaluate 
the 90$\%$ confidence-level Bayesian upper limit branching fraction ($\mathcal{B}_{\textit{UL}}$). This upper limit is obtained by integrating the likelihood function:
\begin{equation}
\int^{\mathcal{B}_{\textit{UL}}}_{0}\mathcal{L}(\mathcal{B})d\mathcal{B} = 0.9\int^{1}_{0}\mathcal{L}(\mathcal{B})d\mathcal{B},
\end{equation}
where $\mathcal{L}(\mathcal{B})$ denotes the likelihood value.
The systematic uncertainties are taken into account by replacing $\mathcal{L}(\mathcal{B})$ with a 
smeared likelihood function.
We thus determine the upper limit on the branching 
fraction of $\mathcal{B}(B^{0} \to p \bar{\Lambda} \pi^{-} \gamma)$ to be $6.5 \times 10^{-7}$ 
at the 90\% confidence level.

In conclusion, we have performed a search for $B^{0} \to p \bar{\Lambda} \pi^{-} \gamma$, which proceeds via the $b \to s \gamma$ radiative penguin process,
by using the full $\Upsilon(4S)$ data sample of $772 \times 10^{6}~B\bar{B}$ pairs collected by Belle. No significant signal yield is found and we set the upper limit on the branching fraction to be $6.5 \times 10^{-7}$ at the 90\% confidence level. We also conclude that the decay under study does not follow the expected hierarchy; instead, we find $ {\mathcal B} ( B^0 \to p \bar{\Lambda}\pi^- \gamma) < {\mathcal B} ( B^+ \to  p \bar{\Lambda} \gamma)$.

We thank the KEKB group for the excellent operation of the
accelerator; the KEK cryogenics group for the efficient
operation of the solenoid; and the KEK computer group,
the National Institute of Informatics, and the 
PNNL/EMSL computing group for valuable computing
and SINET4 network support.  We acknowledge support from
the Ministry of Education, Culture, Sports, Science, and
Technology (MEXT) of Japan, the Japan Society for the 
Promotion of Science (JSPS), and the Tau-Lepton Physics 
Research Center of Nagoya University; 
the Australian Research Council and the Australian 
Department of Industry, Innovation, Science and Research;
Austrian Science Fund under Grant No. P 22742-N16;
the National Natural Science Foundation of China under contract 
No.~10575109, 10775142, 10825524, 10875115, 10935008 and 11175187; 
the Ministry of Education, Youth and Sports of the Czech 
Republic under contract No.~MSM0021620859;
the Carl Zeiss Foundation, the Deutsche Forschungsgemeinschaft
and the VolkswagenStiftung;
the Department of Science and Technology of India; 
the Istituto Nazionale di Fisica Nucleare of Italy; 
The BK21 and WCU program of the Ministry Education Science and
Technology, National Research Foundation of Korea Grant No.\ 
2010-0021174, 2011-0029457, 2012-0008143, 2012R1A1A2008330,
BRL program under NRF Grant No. KRF-2011-0020333,
and GSDC of the Korea Institute of Science and Technology Information;
the Polish Ministry of Science and Higher Education and 
the National Science Center;
the Ministry of Education and Science of the Russian
Federation and the Russian Federal Agency for Atomic Energy;
the Slovenian Research Agency;
the Basque Foundation for Science (IKERBASQUE) and the UPV/EHU under 
program UFI 11/55;
the Swiss National Science Foundation; the National Science Council
and the Ministry of Education of Taiwan; and the U.S.\
Department of Energy and the National Science Foundation.
This work is supported by a Grant-in-Aid from MEXT for 
Science Research in a Priority Area (``New Development of 
Flavor Physics''), and from JSPS for Creative Scientific 
Research (``Evolution of Tau-lepton Physics'').

\end{document}